\documentclass{aastex61}

\newcommand\aastex{AAS\TeX}

\newcommand{\Lyy}{Ly-{$\alpha$}}
\newcommand{\Ly}{Ly-{$\alpha \;$}}
\newcommand{\A}{\AA$\;$}

\shorttitle{\aastex\ LSR J1835+3259}
\shortauthors{Saur et al.}

\begin{document}

\title{The UV spectrum of the
  Ultracool Dwarf LSR
  J1835+3259 observed with the Hubble Space Telescope 
}

\correspondingauthor{Saur}
\email{saur@geo.uni-koeln.de}

\author[0000-0003-1413-1231]{Joachim Saur}
\affil{Institut f\"ur Geophysik und Meteorologie\\
Universit\"at zu K\"oln \\
Albertus-Magnus-Platz \\
Cologne, 50923, Germany}

\author{Christian Fischer}
\affil{Institut f\"ur Geophysik und Meteorologie\\
Universit\"at zu K\"oln \\
Albertus-Magnus-Platz \\
Cologne, 50923, Germany}

\author{Alexandre Wennmacher}
\affil{Institut f\"ur Geophysik und Meteorologie\\
Universit\"at zu K\"oln \\
Albertus-Magnus-Platz \\
Cologne, 50923, Germany}

\author{Paul D. Feldman}
\affil{Department of Physics and Astronomy\\
The Johns Hopkins University\\
Baltimore \\
MD, USA}

\author{Lorenz Roth}
\affil{School of Electrical Engineering\\
Royal Institute of Technology KTH\\
Stockholm \\
Sweden}

\author{Darrell F. Strobel}
\affil{ Department of Earth and
  Planetary Sciences, and Department of Physics and Astronomy\\
The Johns Hopkins University\\
Baltimore \\
MD, USA}

\author{Ansgar Reiners}
\affil{Institut f\"ur Astrophysik \\
Georg-August-Universit\"at\\
Friedrich-Hund-Platz 1 \\
37077 G\"ottingen \\
Germany}



\begin{abstract}

An interesting question about ultracool dwarfs recently raised in
the literature is whether their emission is
purely internally driven or partially powered by external processes
similar to planetary aurora known from the solar system. 
 In this work we present Hubble Space Telescope
observations of the energy fluxes of the M8.5
ultracool dwarf LSR J1835+3259 throughout the UV. The obtained spectra
reveal that the object is generally UV-fainter compared to other earlier-type dwarfs. 
We  detect the \ion{Mg}{2} doublet
at 2800 \A and constrain an average
flux throughout the Near-UV.
 In the Far-UV without  Lyman alpha,
the ultracool dwarf is extremely
faint with an energy output  at least a factor of 1000 smaller as
expected from auroral emission physically similar to that on Jupiter.  
We also detect the red wing of the 
Lyman alpha emission. 
Our overall finding is that the observed UV spectrum of LSR J1835+3259
resembles the spectrum of mid/late-type M-dwarf stars relatively well, but it
is distinct from a spectrum expected from Jupiter-like auroral processes.

%
%
%
\end{abstract}

\keywords{
stars: individual LSR J1835+3259 --- ultraviolett: stars --- stars: low-mass --- brown dwarfs}



\section{INTRODUCTION} \label{sec:intro}
LSR J1835 + 3259 is an ultracool dwarf of spectral type M8.5, which is located
5.6 pc away from Earth in the constellation Lyra
\citep{reid03,lepi03,hall08,desh12}. 
In the Hertzsprung-Russell diagram it is positioned near the end of
the main sequence. In that region the X-ray energy, indicative for the
presence of a magnetically heated corona, drops by two orders of
magnitude over a small range in spectral type
\citep{hall15}. 
LSR J1835 + 3259 is a
fast rotator with a period of 2.84 hours.
Its emission has been observed to be periodic at radio wavelengths, in optical bands and the H-$\alpha$
\citep{hall08,berg08,hall15}.

The dwarf LSR J1835+3259  possesses a strong
magnetic field with values of $\sim$0.2 Tesla derived from radio observations \citep{hall15}.
Observations of NIR polarized and optical emission with the Keck
telescope based on Zeeman signatures demonstrated that its 
magnetic field is at least 0.51 Tesla and covers
at least 11\%
of its visible surface  \citep{berd17,kuzm17}.  
These authors derive from the Keck observations
an effective temperature of $T_{eff}=2800 \pm
30$ K and log gravity acceleration of $log\; g=4.50 \pm 0.05$. Based on
evolutionary models, \cite{berd17} infer a mass of $M = 55 \pm 4$
$M_J$, a radius of $R = 2.1 \pm 0.1 R_J$, and an age $t = 22 \pm
4$ Myr (with $M_J$ and $R_J$ the mass and radius of Jupiter, respectively).
Therefore, \cite{berd17} and \cite{kuzm17} conclude that LSR
J1835+3259 is
a young brown dwarf at the end of its accretion phase. 
We note, however, that at this young age, brown dwarfs exhibit a
strong spectroscopic absorption feature from Li at 6708\,\AA. Brown
dwarfs need a few 10--100 Myr to deplete Li, which makes the Li test a
strong indicator for the age of a young object \citep{basr00}. In
LSR~J1835+3259 no significant absorption of Li was found by
\citet{rein09} in their high-resolution spectra, which appears to be
inconsistent with the parameters derived by
\citet{berd17}. Thus, the brown dwarf status of LSR J1835+3259  is
inconclusive.  In this work, the object is therefore referred to as ultracool
dwarf. They are defined by spectral type M7 and later and thus can
include stars at the end of the main sequence and substellar brown
dwarfs \cite[see
e.g.,][]{rein09,pine17}. 
However, the mass of LSR J1835+3259 is of little importance for our
study although a determination of its age would be extremely useful
for our understanding of low mass object evolution.

Low mass stars including ultracool dwarfs have been extensively studied throughout the
electromagnetic wavelength range including many studies at various UV
wavelengths \cite[e.g.,][]{hawl03,walk08,walk09,fran12,shko14}.  
These objects can 
experience 
persistent
internal heating of their coronal plasma to over 10$^6$
K. This heating above the photosphere is often thought to occur
through dissipation of wave energy and is 
related to strong and localized magnetic
field structures, such as loops \citep[e.g.,][]{pine17}.
Late-type M-dwarfs 
have also been shown to display
flare activity, 
which additionally can contribute to the coronal heating
\citep[e.g.,][]{hall08,rein08,rein09,rein10}. 
Very recently \Ly has been detected on the very late-type M8 dwarf star
Trappist-1 by \cite{bour17,bour17a}. Trappist-1 is the coldest M dwarf
star with  known \Ly emission. It possesses an effective temperature of 2550
$\pm55$ K \cite[]{gill16} with a \Ly flux of 0.05 erg s$^{-1}$
cm$^{-2}$ at 1 AU from the star \cite[]{bour17}. For confirmed brown
dwarfs, no Far-UV emission has been detected, yet \cite[]{pine17}. 
UV observations of brown dwarfs would be 
highly interesting because brown dwarfs possess properties in between
those of low mass
dwarf stars and very massive planets.  They are massive enough to burn deuterium or
lithium, but not massive enough to fuse hydrogen
\citep[e.g.,][]{basr14}.  The atmospheres of brown dwarfs 
also display planet-like weather
phenomena and clouds including species such as TiO 
\citep{cros14,hell14}. 

An exciting new possibility about the ultracool dwarf LSR J1835+3259  was raised by
\cite{hall15}. The authors  reported simultaneous radio and optical 
emissions from LSR J1835+3259, which was interpreted as auroral
emission triggered by electron beams generated outside of the 
dwarf.  This would be the first auroral emission observed outside the
Solar System, but a confirmation of the proposed hypothesis is
crucial. \cite{hall15} argue that the auroral emission on the dwarf
could be powered by processes similar to Jupiter, i.e., 
by magnetospheric currents that couple energy into the
upper atmosphere. 
 The nature of this emission  would then be fundamentally different
 compared to 
the emission from stars like the Sun, which is powered by internal
processes, e.g., by magnetic processes
that occur in their convection zones.  
The conclusions in
\cite{hall15}  have
been reached by detailed modeling of the observed light
curves. The derived power emitted at radio wavelength is 
10$^{22}$ erg s$^{-1}$,
 which
requires 
10$^{24}$ to 10$^{26}$ erg s$^{-1}$
of available power in electron beams. The
Balmer H-$\alpha$ line emission amounts to a total of 
$2.5 \times 10^{24}$ erg s$^{-1}$.
No X-Ray
emission associated with the presence of a magnetically heated corona
has been observed on LSR J1835 + 3259 by the Chandra Observatory
\citep{berg08}. Near/Mid-ultraviolet emissions around 2600
\A with FWHM of 693 \A have marginally been detected with the SWIFT 30 cm
telescope \citep{berg08}. The integrated NUV
flux however still amounts to a total of 
$2.5 \times 10^{25}$ erg s$^{-1}$.

Auroral phenomena and electromagnetic coupling in extrasolar
magnetospheres and astrospheres is a topic of growing interest from the observational side 
\citep[e.g.,][]{shko03,shko08,popp11,fran13,pine17} and the
theoretical side
\citep[e.g.,][]{cunt00,ip04,preu05,lanz09,cohe09,saur13,stru15,saur17}. Whereas
this body of work refers to the more general auroral and electromagnetic 
coupling of planets and stars,
auroral and related radio emission from brown dwarfs and ultracool 
stars specifically have been studied theoretically by, e.g., 
\cite{schr09,nich12,turn17}. In the latter two studies it has been assumed that the
steady-state-current picture of the auroral processes of Jupiter can
be carried  over to the dwarfs.

Observations and modeling of auroral processes and the associated electromagnetic
couplings have a long tradition in solar system research.
From the knowledge in the solar system, aurorae need three components
\citep[e.g.,][]{mauk12}: (1) A “generator”, located in the
planet’s magnetosphere, which produces electric current or more
generally electromagnetic energy. The electric
current and energy are continued along the object's magnetic field lines into the
atmosphere/ionosphere. (2) An “accelerator region” energizes the low
energy electrons (which in most cases carry the current) to energies
up to keV or MeV energies. (3) The electrons in the form of beams
finally precipitate onto the atmosphere where atoms and molecules
serve as the “screen” from which emission over a broad spectral range
spanning UV, visible to IR and radio wavelengths are excited by
electron impact. Observations of the relative ratios of the emission
intensities at these wavelengths and their temporal/spatial structure
give essential insights into the three auroral components and thus
into the plasma environment around planetary objects.

In the solar system, Jupiter is the most
massive planet with the strongest magnetic field, which also 
possesses the solar system's most
powerful aurora. Compared to LSR J1835 + 3259, Jupiter's magnetic
field is weaker with a polar field of $\sim10^{-3}$ T and Jupiter is a slower
rotator with a period of 9.9 hours. Aurora on Jupiter has been extensively
studied spatially and at all wavelengths with the following
characteristic total energy fluxes  \citep{bhar00}: 
X-Ray: 1-4 $ \times 10^{16}$ 
erg s$^{-1}$, 
 Far-UV (800-1800 \AA): $2-10 \times 10^{19}$ 
erg s$^{-1}$, 
visible
(3850-10000 \A): $10-100 \times 10^{16}$ 
erg s$^{-1}$, 
IR: $50 \times 10^{19}$ 
erg s$^{-1}$ 
 and radio (10kHz to a
few MHz): $10 \times 10^{16}$ 
erg s$^{-1}$. 
Most of the energy in the visible is within
the Balmer lines. The origin of the X-ray emission is not fully
understood, but is thought to be generated by different
processes compared to the aurora at other wavelengths
\citep{glad02}. The \Ly spectrum of Jupiter has been studied in
detail by a series of authors \citep[e.g.,][]{clar89,clar94a,pran97}. 
Jupiter's radio emission is caused by the electron cyclotron maser
instability generated by its non-thermal, auroral electron
distributions \cite[e.g.,][]{zark98}.

Jupiter's aurora has three qualitatively different spatial features. (1):
The main auroral oval is generated by the breakdown of magnetospheric
corotation, when the plasma originating from Jupiter's moon Io moves
radially outward  \citep[e.g.,][]{hill01,clar02}. 
This establishes a process, similar to magnetic braking at stars,
which couples Jupiter to its magnetospheric plasma.
Therefore  Jupiter's rotational
energy ultimately powers
the emission of its main auroral oval. (2) The moons of
Jupiter leave auroral imprints in Jupiter's atmosphere
\citep[e.g.,][]{conn93,clar02,wann10,bonf17,saur13}. (3): Polar
emission is the least understood and likely originates
from  the outermost region of its magnetosphere.
Our understanding of Jupiter's aurora
currently experiences a paradigm change with the NASA spacecraft JUNO in a
polar orbit around Jupiter. Among the new findings are that the dominant part of the auroral
electrons appears to be broad in their energy distribution and are accelerated towards
and away from Jupiter simultaneously \citep[e.g.,][]{mauk17,eber17}. This
points to the importance of stochastic acceleration  potentially 
powered by plasma waves as, e.g., evoked in
\cite{saur03} compared to
acceleration processes only related to large scale steady-state
magnetospheric
current systems.

In this work, we 
analyze
the UV spectrum of the 
dwarf LSR J1835+3259 to further investigate its possible auroral emission.
Therefore we will test whether  the spectral energy density (SED) of LSR J1835+3259
scales similar to the auroral emission from Jupiter or similar to the emission
from mid/late-type M stars.
%
%
%
In section \ref{sec:obs}, we first present details of new observations
 by the Space Telescope Imaging Spectrograph (STIS) on the Hubble Space Telescope (HST)
 to characterize the spectrum of LSR J1835+3259 throughout the
UV. We also detail our data analysis procedures to search for the very
faint emission of the target within these observations. In
section \ref{sec:results}, we present and discuss the observed fluxes for various
wavelength ranges. In
section \ref{sec:implications} we compare our results with the spectral properties
of Jupiter and M-dwarfs  and in section \ref{sec:conclusions} we
discuss our main findings.

\section{OBSERVATIONS} \label{sec:obs}

The HST/STIS observations of  LSR J1835+3259 (program  ID 14617) are
designed to search for auroral emission throughout the UV wavelength range. 
In Table \ref{tab:obs}, we summarize the details of the 9 STIS 
exposures taken during 
five consecutive orbits of HST. Orbit number 1, 4, 5 were dedicated to the
Far-UV (FUV) emission, orbit 2 to the Near-UV (NUV) and orbit 3 to the \Ly wings. 
\begin{deluxetable*}{ccccccccccc}[b!]
\tablecaption{Exposure details of HST/STIS observations of LSR J1835+3259
  (ID: 14617)\label{tab:obs}}
\tablecolumns{9}
\tablenum{1}
\tablewidth{0pt}
\tablehead{
\colhead{Orbit} 
&\colhead{Exp } 
& \colhead{Rootname} 
&\colhead{UT obs date\tablenotemark{a}} 
&\colhead{UT obs time\tablenotemark{a}}
&\colhead{Exp time} 
&\colhead{Type} 
& \colhead{Grating} 
& \colhead{Disp} 
& \colhead{Slit width} 
\\
\colhead{\#}  
&\colhead{\#}  
&\colhead{ }  
&\colhead{yyyy-mm-dd}
& \colhead{ hh:mm:ss} 
&sec
&
&
&\colhead{\AA/pixel} 
& \colhead{arcsec} 
& \colhead{}
}
\startdata
1&1&od9x01010 &2017-03-04 & 15:59:14  & 1083.2 & FUV               & G140L  & 0.584 & 0.2 \\
1&2&od9x01020 &2017-03-04 & 16:20:43  & 1084.2 & FUV               & G140L  & 0.584 & 0.2 \\
2&3&od9x01030 &2017-03-04 & 17:21:22  & 1395.2 & NUV              & G230L  & 1.548 & 0.2 \\
2&4&od9x01040 &2017-03-04 & 17:48:03  & 1394.2 & NUV              & G230L  & 1.548 & 0.2 \\
3&5&od9x01050 &2017-03-04 & 19:13:04  & 2020.2 & Ly-$\alpha$ & G140M  & 0.053 & 0.05 \\
4&6&od9x01060 &2017-03-04 & 20:32:16  & 1395.2 & FUV               & G140L  & 0.584 & 0.2 \\
4&7&od9x01070 &2017-03-04 & 20:58:57  & 1394.2 & FUV               & G140L  & 0.584 & 0.2 \\
5&8&od9x01080 &2017-03-04 & 22:07:43  & 1395.2 & FUV               & G140L  & 0.584 & 0.2 \\
5&9&od9x01090 &2017-03-04 & 22:34:24  & 1394.2 & FUV               & G140L  & 0.584 & 0.2 \\
\enddata
\tablenotetext{a}{At exposure start.}
\end{deluxetable*}


The data analysis in this work is based on the x2d files where
spectral energy
fluxes in erg s$^{-1}$ cm$^{-2}$ \AA$^{-1}$ are provided.
Because of
the extremely faint nature of the target in the UV, 
its position on the detector and in the resultant x2d files could not
be determined by direct  visual identification.
To search for and to extract fluxes, we
applied the following procedure. The nominal reference location of the target
in the $y$ direction is $y_{ \mbox{ref,nom}}$. The associated row is calculated
by $y_r=\mbox{integer}(y_ {\mbox{ref,nom}}+0.5)$. The x direction in the x2d files is
the direction of dispersion, i.e., the direction of wavelength $\lambda$. We
extract the fluxes $f_{trace}$ in $x$ direction along the trace within several rows
above and below the
reference location $y_r$. The flux per unit
wavelength as function of
column $i_x$ or equivalently $\lambda$ is given by
\begin{eqnarray} 
f_{trace}(\lambda)  =\sum_{i_y=y_{ref}- n_1
}^{y_{ref}+ n_2 }  f(\lambda, i_y) \;. 
\end{eqnarray}
The average background flux per pixel $f^{px}_{bg}$ is
calculated from rows sufficiently above and below the rows where we expect
flux from the target, i.e.,
\begin{eqnarray} 
f^{px}_{bg}(\lambda) =
\left(
\sum_{i_y=y_{ref}+ a_1
}^{y_{ref}+ a_2 }  f(\lambda, i_y)  + \sum_{i_y=y_{ref}- b_1
}^{y_{ref}- b_2 }  f(\lambda, i_y)
\right)  / (a_2-a_1+1+b_2-b_1+1)
\end{eqnarray} 
with the positive integer numbers $n_1, n_2, a_1, a_2,
b_1, b_2$. The value of these integers depend on the 
type of exposure (see Table \ref{tab:data_analysis}). This procedure
calculates 
a separate background for
each column $i_x$. The reason is that the
background fluxes change along the dispersion axis.
We assume that 
the background flux 
characterizes the background along the trace and thus
the net flux from the target as a function of wavelength is given by
\begin{eqnarray}
f_{net}(\lambda) = f_{trace}(\lambda) - 
f^{px}_{bg}(\lambda)\left(
n_2-n_1+1
\right)
\;.
\end{eqnarray}

The variance  of the background for each pixel in an individual column
$i_x$ associated with a certain wavelength $\lambda$
is given by
\begin{eqnarray}
V^{px}(i_x)=
\left(
\sum_{y_{ref}+ a_1
}^{y_{ref}+a_2 }  (f(i_x, i_y) -f^{px}_{bg}(i_x))^2+ \sum_{y_{ref}- b_1
}^{y_{ref}-b_2 }  (f(i_x, i_y) -f^{px}_{bg}(i_x))^2 
\right)
/  (a_2-a_1+b_2-b_1+1) 
\;.
\end{eqnarray}
When we compare flux uncertainties with the net fluxes from the target
$f_{net}$,  the variance in each pixel $V^{px}$ needs to be normalized to the number of rows,
which have been used to calculate $f_{net}$. In case net  fluxes within 
certain wavelength ranges are calculated, the variances of the
individual columns, which contribute to the selected wavelength
ranges, additionally need to be summarized to the total variance $V$. 
The resultant uncertainties are calculated based on
the standard deviation $\sigma=\sqrt{V}$.

\begin{deluxetable*}{cccCCCCc}[b!]
\tablecaption{Data analysis details: Row information used for
  calculation of net fluxes. \label{tab:data_analysis}}
\tablecolumns{7}
\tablenum{2}
\tablewidth{0pt}
\tablehead{
\colhead{Type} 
&\colhead{$y_r$}
&\colhead{$n_1$}
&\colhead{$n_2$}
& \colhead{$a_1$}
& \colhead{$a_2$}
& \colhead{$b_1$}
& \colhead{$b_2$}
}
\startdata
NUV & 601& -3&  2  &  3 & 23&-24&-4 \\
FUV  & 474& -2& 4   &  5 & 45&-43&-3 \\
\Ly & 488&  -2&  4  &  5 & 35&-33&-3  \\
\enddata
\end{deluxetable*}

\section{RESULTS} \label{sec:results}
The results of our observations are now discussed separately for the
three different types of observation, i.e., 
for the NUV wavelength range
in subsection \ref{subsec:NUV}, the FUV wavelength range in subsection
\ref{subsec:FUV} and the \Ly
wavelength in subsection \ref{subsec:Ly}.

\subsection{Near-UV} \label{subsec:NUV}
\begin{figure}
\plotone{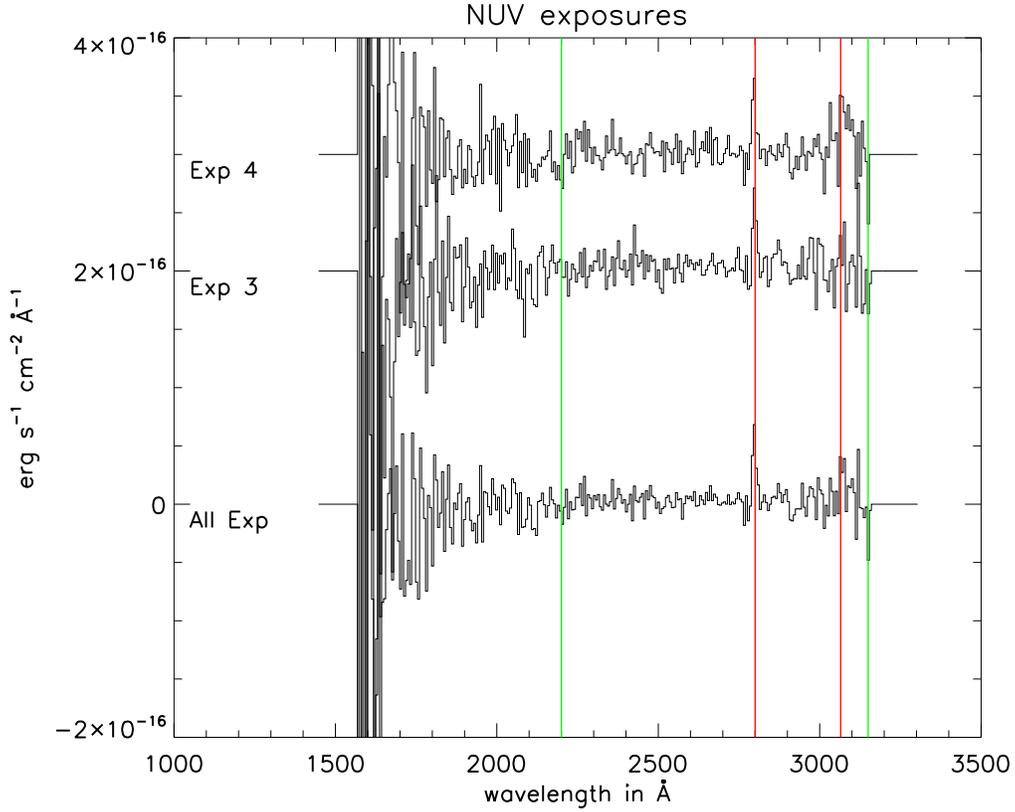}
\caption{Net NUV flux per wavelength along trace for exposure \# 3 and \# 4
  and combined exposure. Spectra are shifted for readability. The green vertical lines
  mark the wavelength range 2200 \A to 3150 \A over which we
  calculate the total FUV flux. The red vertical line at 2800 \A
  indicates the expected wavelength of \ion{Mg}{2} emission and the red line at 3065
  \A might be associated with TiO emission (see main text). \label{fig:NUV-TRACE}
}
\end{figure}

NUV observations were performed with STIS grating G230L within
the wavelength range 1570  - 3180 \AA. Observations were taken during
orbit 2 with two exposures (see Exp \# 3 and \# 4 in
Table \ref{tab:obs}). The net fluxes per \A along the trace for each exposure and
the combined NUV exposures are shown in Figure
\ref{fig:NUV-TRACE}. 
The net flux $f_{net}(\lambda)$ is calculated as described in section \ref{sec:obs}.
In order to minimize the contribution of the background noise, the numbers of rows to
calculate the net flux was kept at a relatively small value, i.e., 
6
(see
Table \ref{tab:data_analysis}). 

\begin{figure}
\plotone{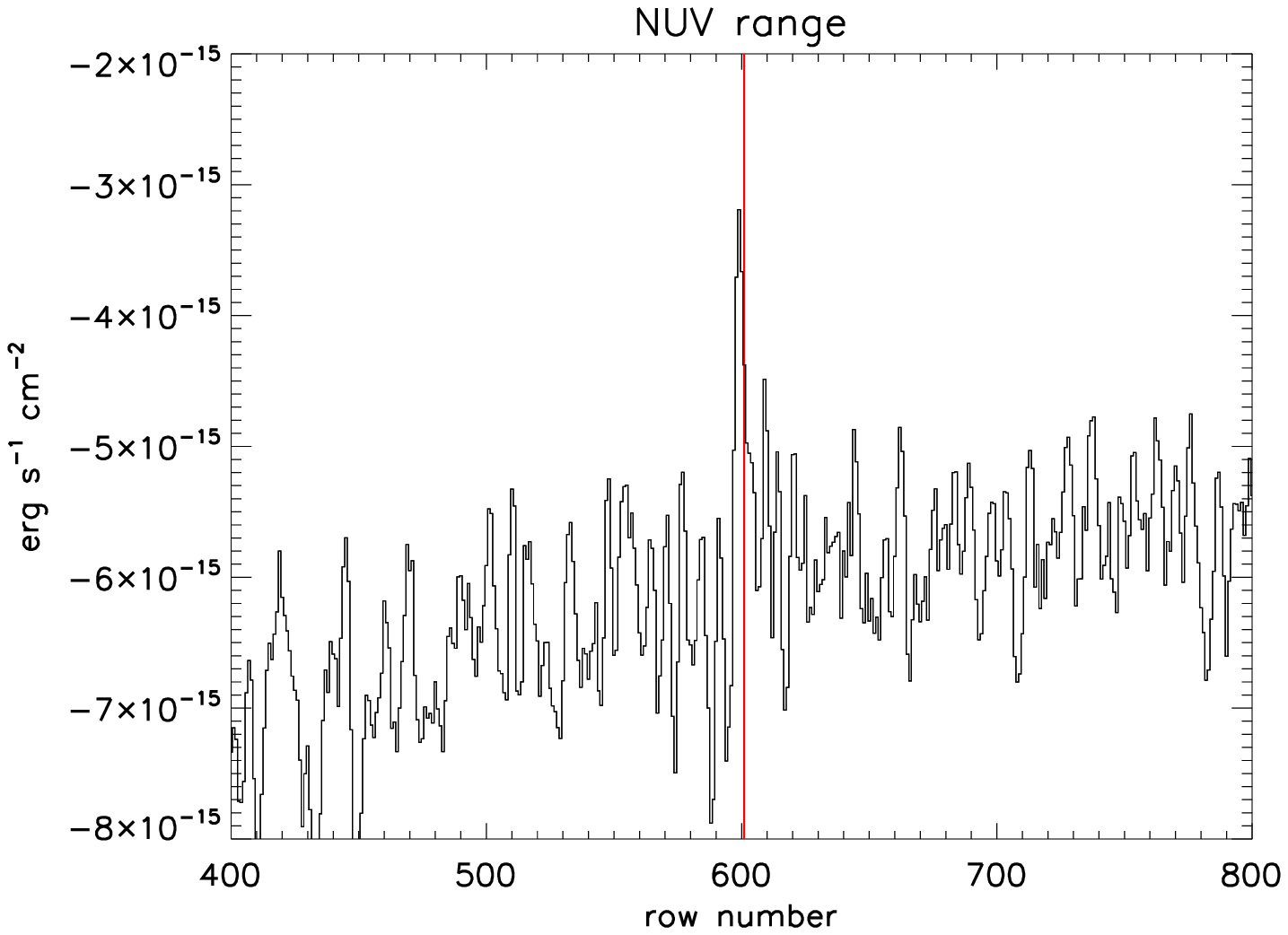}
\caption{NUV flux integrated along the direction of dispersion as a
  function of the row number i$_y$ to confirm location of target. The red
  vertical line 
shows the expected
  reference location of LSR J1835+3259.  Note that the negative values
  of the background fluxes occurring in the x2d-files
do not have physical meaning. 
They are due to the dark current correction within the
STIS calibration pipeline of STScI generating the x2d-files. A detailed discussion of the dark current uncertainties 
in the NUV MAMA data can be found in section 4.1.3 of the STIS Data
Handbook
\cite[]{bost11}.
Only for the purpose of target confirmation
within the background noise, we show here the original x2d data and do not remove
these background fluxes in this figure.
\label{fig:NUV-INT}}
\end{figure}

Due to the faint nature of the object in the UV, we 
independently check whether the detected NUV flux is indeed
collocated with the target position. Therefore
we display in
Figure \ref{fig:NUV-INT} the flux 
integrated along the dispersion direction between 2200 \A and 3015 \A as
a function of row number
where we added the neighboring row above and the one below,
respectively, in order to reduce the scattering in the plot. 
The reference location of the target 
lies at row 601 indicated by the red vertical line. The maximum of
the integrated flux approximately coincides with the reference location and thus
confirms 
a detection of
a surplus of NUV flux generated by LSR J1835+3259. 
The reference is  located one row number higher
compared to the maximum in Figure \ref{fig:NUV-INT}, which we adjust
by our choice of coefficients in Table \ref{tab:data_analysis} when
calculating spectra.
A visual comparison with
the integrated flux values of the
neighboring traces shows that the
surplus along the reference trace is  roughly 3 - 4 $\times
10^{-15}$ erg s$^{-1}$ cm$^{-2}$.

When we integrate the net fluxes displayed in Figure \ref{fig:NUV-TRACE} between 2220 and 3150 \AA, we find
a total averaged flux from both exposures of (3.9 $\pm$0.58) $\times 10^{-15}$ erg s$^{-1}$ cm$^{-2}$ (see also
Table \ref{tab:energy_fluxes}). 
This leads to an average flux per wavelength of  (4.1 $\pm$0.61) $\times 10^{-18}$ erg s$^{-1}$ cm$^{-2}$
\AA$^{-1}$. This value is similar to that from previous observations of ($4.7 \pm 1.3$)
$\times 10^{-18}$ erg s$^{-1}$ cm$^{-2}$ \AA$^{-1}$ 
obtained with the SWIFT Telescope  in the NUV 
\cite[]{berg08}. 
\begin{deluxetable*}{llCCCRC}[b!]
\tablecaption{Observed energy fluxes\label{tab:energy_fluxes}}
\tablecolumns{7}
\tablenum{3}
\tablewidth{0pt}
\tablehead{
\colhead{Exposures} 
&\colhead{Properties}
&\colhead{Wavelength}
& \colhead{Energy flux}
& \colhead{S/N}
& \colhead{Spectral energy flux}
& \colhead{}
\\
\colhead{}
& \colhead{}  
& \colhead{\A}  
& \colhead{erg s$^{-1}$ cm$^{-2}$}  
& \colhead{ }  
& \colhead{erg s$^{-1}$ cm$^{-2}$} \A$^{-1}$  
& \colhead{}    
}
\startdata
Exp 3 & NUV:  \ion{Mg}{2}& 2792.0 - 2807.0&  (9.3 \pm 0.57) $\times$ 10$^{-16}$ & 16.4& (6.2 \pm 0.38) $\times$ 10$^{-17}$ & \\
Exp 4 & NUV:  \ion{Mg}{2}& 2792.0 - 2807.0&   (7.1 \pm 0.70) $\times$10$^{-16}$& 10.1   &(4.7 \pm 0.47) $\times$ 10$^{-17}$   & \\
Exp 3+4 & NUV:  \ion{Mg}{2}& 2792.0 - 2807.0&   (8.2 \pm 0.46) $\times$ 10$^{-16}$ & 18.0 & $ (5.5 \pm 0.31) $\times$ 10$^{-17}$ & \\ 
Exp 3 & NUV & 2200.0 - 2700.0&  (1.7 \pm 0.35) $\times$ 10$^{-15}$ &4.9& (3.4 \pm 0.69) $\times$ 10$^{-18}$ & \\ 
Exp 4 & NUV & 2200.0 - 2700.0&  (1.2 \pm 0.35) $\times$ 10$^{-15}$ & 3.4&(2.4 \pm 0.70) $\times$ 10$^{-18}$ & \\
Exp 3+4 & NUV & 2200.0 - 2700.0&  (1.4 \pm 0.24) $\times$ 10$^{-15}$ &5.9&  (2.9 \pm 0.49) $\times$ 10$^{-18}$ & \\
Exp 3+4 & NUV & 2200.0 - 3150.0&  (3.9 \pm 0.58) $\times$ 10$^{-15}$ &6.7&  $ (4.1 \pm 0.61) $\times$ 10$^{-18}$ & \\
Exp 1+2+6+7+8+9 & Average FUV & 1330.0 - 1710.0&  (1.9 \pm 1.3) $\times$ 10$^{-16}$  &1.4&  (4.9 \pm 3.6) $\times$ 10$^{-19}$  & \\
Exp 5 & \Ly \hspace{0.05cm}  red wing& 1215.85-1216.2 & (8.0 \pm 1.6) $\times$ 10$^{-16}$ &4.9& (2.3 \pm 0.47) $\times$ 10$^{-15}$ \\
\enddata
\end{deluxetable*}

\begin{figure}
\plotone{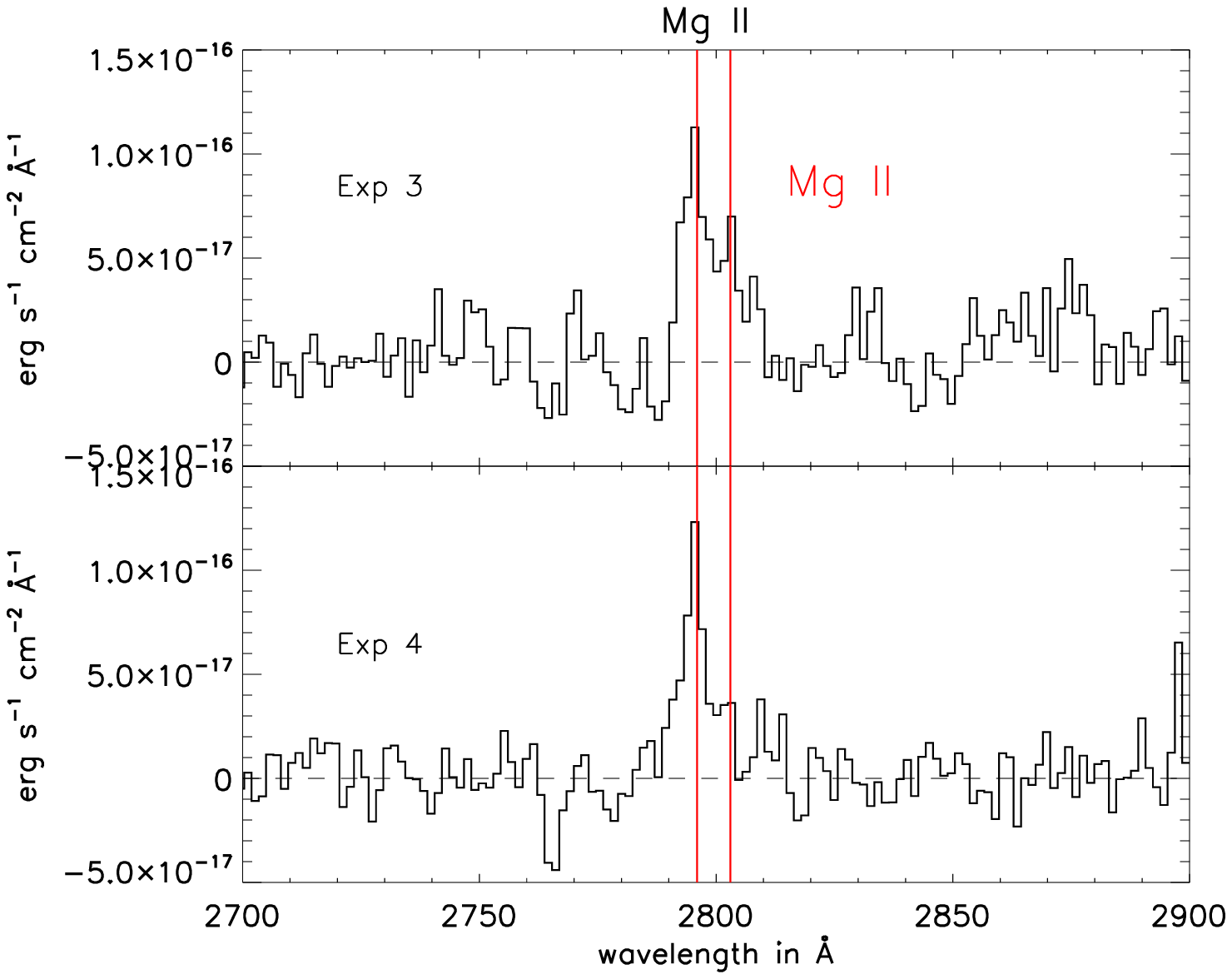}
\caption{\ion{Mg}{2} doublet for exposure \# 3 and \#4. The vertical red lines
  indicate the known position of the \ion{Mg}{2} doublet at 2796 and 2803 \AA \citep[e.g.,][]{feld96,fran13}.  \label{fig:NUV-MgII}}
\end{figure}

In Figure \ref{fig:NUV-TRACE},  a spectral feature around 2800 \A is
apparent in exposures \# 3 and \#  4. In Figure \ref{fig:NUV-MgII}, we
zoom into this spectral region 
and identify the feature as the \ion{Mg}{2} doublet at 2796 and 2803 \A
(e.g., \cite{feld96,fran13}). 
In exposure \#3 the double peak structure is better visible
compared to the exposure \#4. The
total flux within the doublet between 2792 and 2807 \A is $ (9.3 \pm
0.57) \times 10^{-16}$ erg s$^{-1}$ cm$^{-2}$ for exposure \# 3 and 
$(7.1 \pm 0.70) \times 10^{-16}$ erg s$^{-1}$ cm$^{-2}$ for exposure
\# 4. The fluxes measured during the two exposures are different at
the approximately
1-$\sigma$ level.
A possible reason could be
that longitudinal variability combined with the rapid rotation of the
dwarf causes an apparent
time-variability. Time-variability of emission at other wavelengths
related to the dwarfs rotation have also been observed by
\cite{berg08} and \cite{hall08,hall15}.
The spectrum contains significant
flux 
between 2200 and 2700 \A with an averaged value of 
$(1.4 \pm 0.24) \times 10^{-15}$ erg s$^{-1}$ cm$^{-2}$ for both
exposures combined with a 
S/N = 5.9
(see Table \ref{tab:energy_fluxes} and Figure \ref{fig:NUV-INT}).
The flux within this wavelength range is by a factor of 1.4 $\pm$ 0.5 larger for exposure \# 3
compared to exposure \# 4. This ratio also points towards a
time/longitudinal variability of the emission 
between both exposures. 
Due to the noisy data, it is however not possible to identify
  spectral features such as 
\ion{Fe}{2} emission within this wavelength band.

In Figure  \ref{fig:NUV-TRACE}, a spectral feature around 3065 \A is
visible, which is marked with a red vertical line. The fluxes within the wavelength band 3060 -- 3080 \A are
significant with an S/N of approximately four. This feature around 3065 \A
might be due TiO emission.  According to \cite{palm72}, TiO possesses the
most prominent features in the UV in a collection of bands within
2900-3260 \AA.  In laboratory experiments,   \cite{path70} identify
unclassified electronic transitions near 3062 \A and 2069 \AA, which \cite{palm72}
measure at a slightly different wavelength of 3071 \AA.  TiO emission from LSR
J1835+3259 would not be very surprising because TiO 
has been shown to be present 
 on M-dwarfs \cite[e.g.,][]{bess91,john96}, on brown dwarfs \cite[e.g.,][]{rebo96}
and on hot Jupiters \cite[e.g.,][]{evan16} 
through observations of absorption features.

\subsection{Far-UV} \label{subsec:FUV}

\begin{figure}[ht!]
\plotone{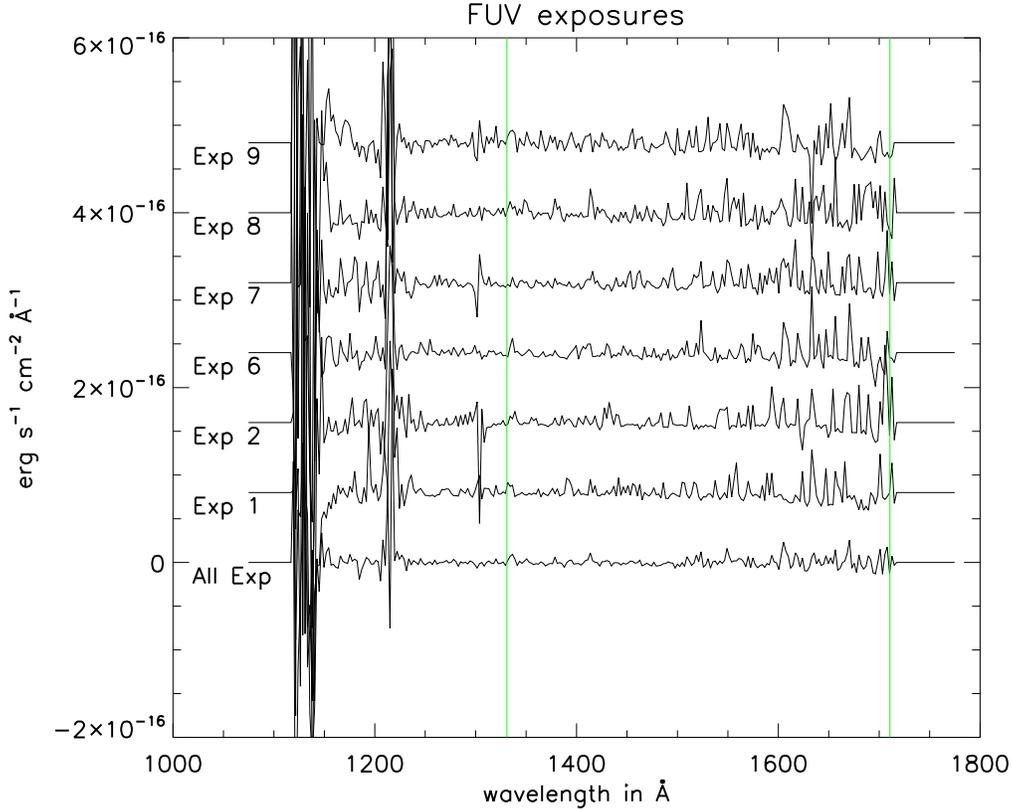}
\caption{Net FUV flux along trace for all FUV exposures and average
  net flux from all FUV exposures. Fluxes for different exposures are
  shifted for visibility.  The green vertical lines
  show the wavelength range over which total fluxes are calculated.
\label{fig:FUV-TRACE}}
\end{figure}

Emission in the FUV is searched for with STIS grating G140L within the
wavelength range 1150 \A to 1730 \AA. Observations were taken during three
visits with the original aim to resolve time-variability of the FUV emission (see Table \ref{tab:obs}).
 In Figure \ref{fig:FUV-TRACE} we show the net fluxes as a
function of wavelength for all six exposures and the average of all exposures.
The spectra are shifted vertically for readability, respectively.
The fluxes contain spectral variability at longer wavelengths. This variability however
appears to be mostly of statistical nature since the average spectrum is
fairly flat. At 1304 \A and \Ly wavelengths the spectrum is strongly contaminated
by geocoronal emission.

\begin{figure}
\plotone{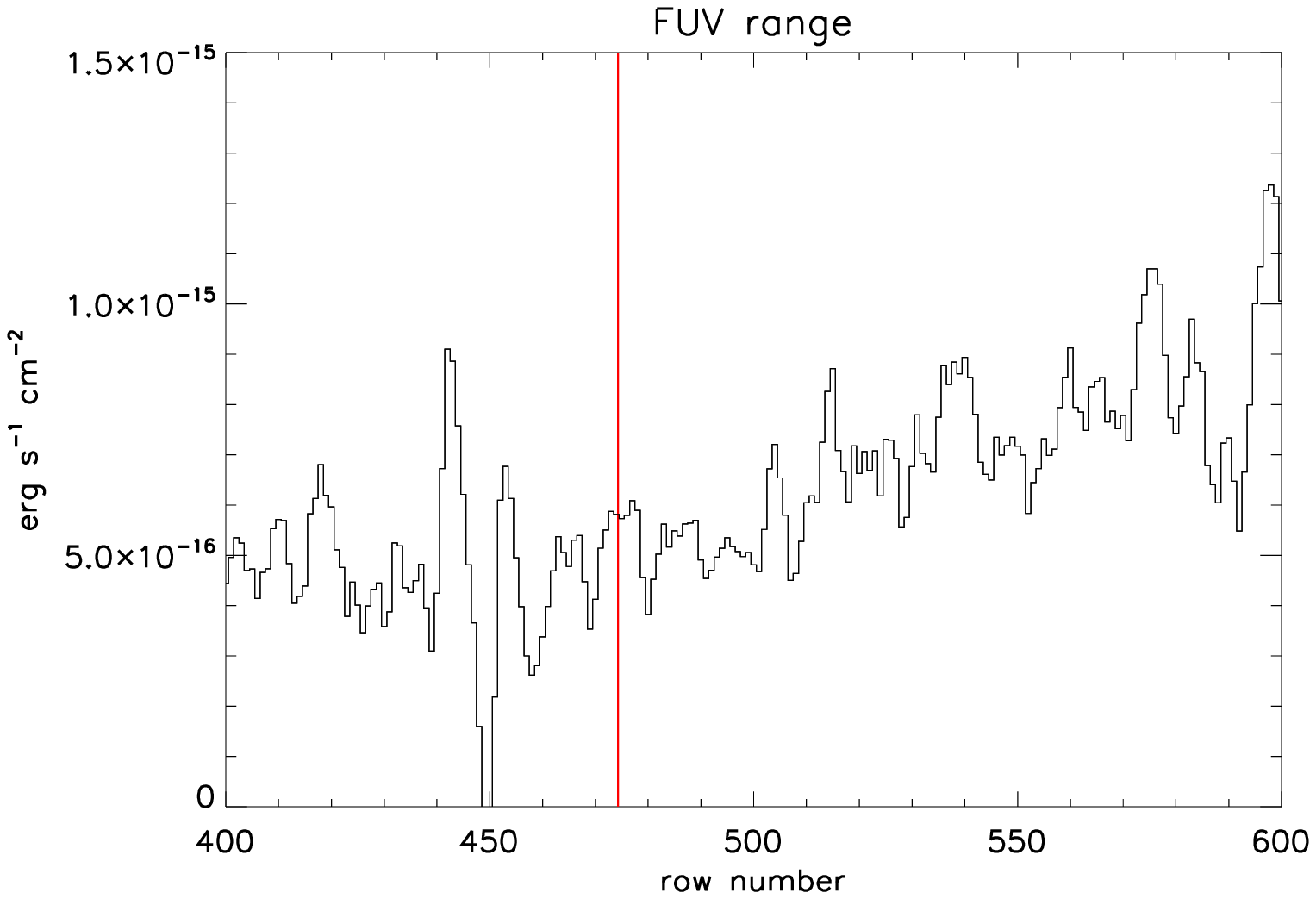}
\caption{FUV flux integrated along direction of dispersion as a
  function of the row number. The red vertical line shows the expected
  reference location of LSR J1835+3259. \label{fig:FUV-INT}}
\end{figure}

Integrating the flux between 1330 \A and 1710 \A indicated by the green
lines in Figure \ref{fig:FUV-TRACE}, we find a total
flux of $ (1.9 \pm 1.3) \times 10^{-16}$  erg s$^{-1}$ cm$^{-2}$. This
flux cannot be considered significant because its uncertainty is
similar to the flux. 
The uncertainty in the detection of a signal from LSR J1835+3259 is
also evident in Figure \ref{fig:FUV-INT}. In this Figure we show the integrated flux  between 1330
\A and 1710 \A  as a function of row number
where we added the flux of the neighboring row above and the one below,
respectively, in order to reduce the scattering in the plot. 
%
The trace where LSR J1835+3259 is formally located is indicated with
a red vertical line,
 which
shows a very small local
maximum. The existence of other larger maxima at neighboring rows
however indicates that no FUV
emission from LSR J1835+3259 can be identified with significance.

\subsection{Lyman-$\alpha$} \label{subsec:Ly}

\begin{figure}
\plotone{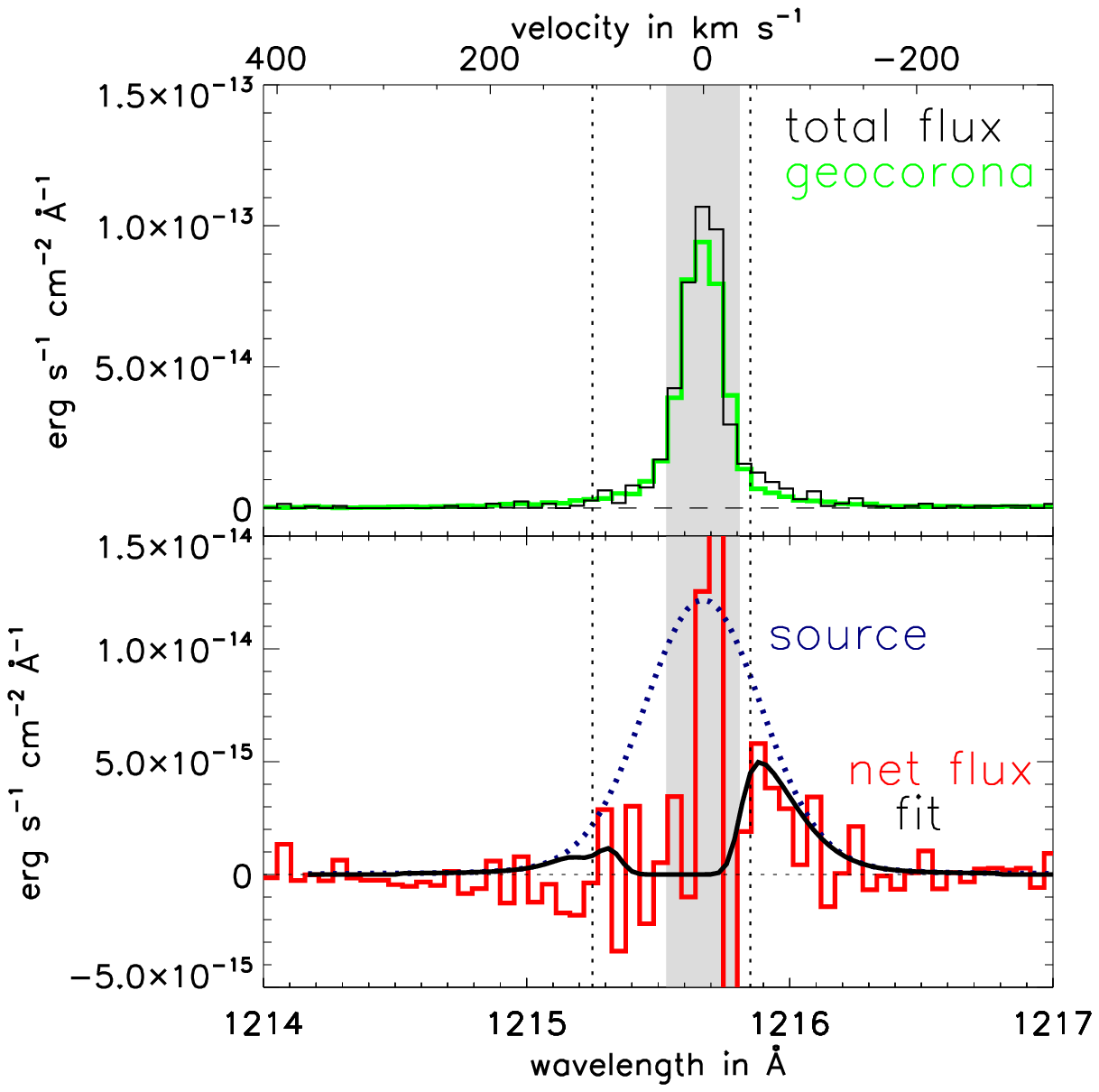}
\caption{\Ly fluxes. Top panel shows total \Ly flux in black and geocoronal
  emission as green line. Bottom panel shows net Ly-$\alpha$ flux in
  red. Wavelengths between the dotted vertical lines indicate
  where the target emission experiences very
  significant absorption in
  the interstellar medium. The blue dotted curve is the reconstructed source \Ly
  profile. The black solid lines is a fit to the data calculated by
  considering absorption of \ion{H}{1}  and \ion{D}{1} in the interstellar
  medium. The grey shaded area shows the wavelength range heavily
  contaminated by geocoronal emission.
\label{fig:LYM-TRACE}}
\end{figure}

During orbit 3, we searched for emission from the \Ly wings of LSR
J1835+3259 using the grating G140M and a very narrow slit of 0.05
arcsec. 
The resultant spectra are shown in Figure \ref{fig:LYM-TRACE}. 
In the top panel, we show in black the total spectrum along the trace,
which includes rows with the target. Emission in the spectral
vicinity of the \Ly resonant  emission line at 1215.67 \A is
strongly altered by absorption in the interstellar medium and by
geocoronal emission \citep[e.g.,][]{lins96,lins14,vida03,fran13}. The 
geocoronal \Ly-emission is displayed 
in Figure \ref{fig:LYM-TRACE} as the green curve. It is calculated from
rows not containing the target (see Table \ref{tab:data_analysis}). 
The emission inside the wavelength
range 1215.35 \A and 1215.85 \AA, 
 indicated as vertical dotted lines in Figure
\ref{fig:LYM-TRACE}, is typically very significantly affected by absorption in the
interstellar medium \citep[e.g.,][]{lins14}. 

The net flux, i.e., the difference between the total flux and the
geocoronal background is displayed in the lower panel of Figure
\ref{fig:LYM-TRACE} as red curve.  In the
red wing around 1216 \A a surplus is visible (both in the top and
bottom panels). There is no significant net flux in the very noisy blue wing. 
The spectrum in the red wing yields values up
$ \sim$5$ \times 10^{-15}$  erg s$^{-1}$ cm$^{-2} $ \AA$^{-1}$ 
near 1215.9 \AA. For the total integrated flux in the red wing between 1215.85 \A
and 1216.2 \A we find $ (8.0 \pm 1.6) \times 10^{-16}$  erg s$^{-1}$
  cm$^{-2}$.

To independently test the significance of the surplus of emission in the
\Ly wings, we compute the total flux within the red wing between 1215.85
\A and 1216.2 \A in each row separately.
\begin{figure}
\plotone{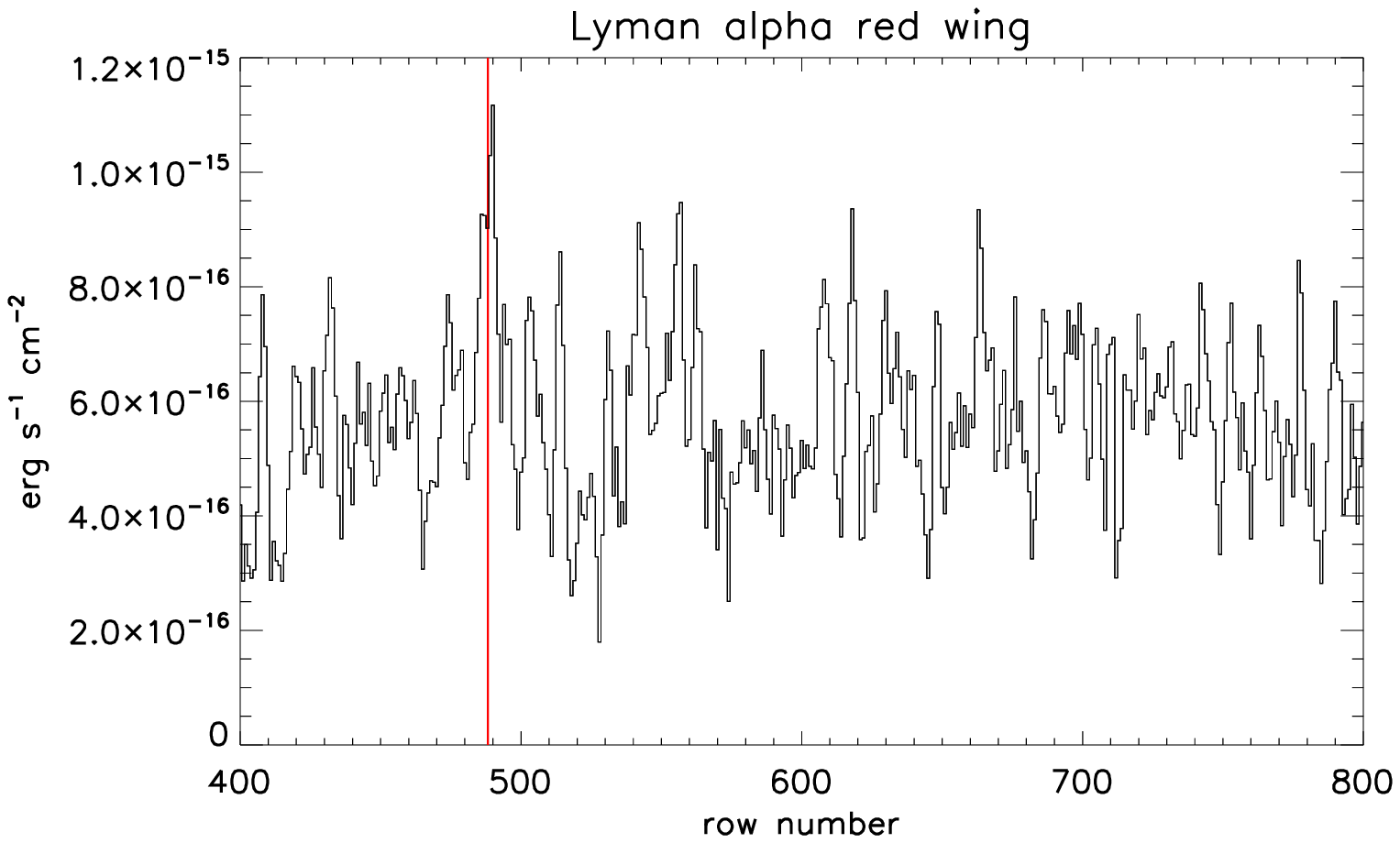}
\caption{Integrated flux within \Ly wing within 1215.85 and 1216.2
  \A as a function of 
 row number. The red vertical line shows the expected
  reference location of LSR J1835+3259. \label{fig:LYM-INT}}
\end{figure}
The resultant integrated flux as a function of row number
 is shown
in Figure \ref{fig:LYM-INT}.
In the integration along a row we added the flux of the neighboring row above and the one below,
respectively, in order to reduce the scattering in the plot. 
 We see a local surplus of
emission at the reference row indicated as red vertical line. The flux at the reference row is
maximum even though smaller local maxima exist at other rows.

Based on the observations of the \Ly fluxes in the red wing, we
reconstruct  the expected \Ly source profile from LSR
J1835+3259 similarly to previously applied approaches
\cite[e.g.,][]{wood05,fran13,bour15,youn16}.
 We model the source profile with a Voigt
profile with a Doppler width of 73 km s$^{-1}$ and a damping
parameter of 0.09 (similar to the values for the M3 star GJ 436 chosen 
by \cite{bour15}).
Absorption in the interstellar medium by hydrogen \ion{H}{1} and deuterium
\ion{D}{1} was calculated with a  Lorentzian absorption profile with cross
sections from, e.g., \cite{mort03} and \cite{wies09},  and a
Maxwell-Boltzmann
velocity distribution of the interstellar gas with 
an adjusted Gaussian standard deviation of
 $\sigma_H = 12 $ km s$^{-1}$ . 
The ratio of \ion{D}{1} to \ion{H}{1}
was assumed to be 1.5 $\times$ 10$^{-5}$ and  standard deviation of
deuterium $\sigma_D$  to be a factor of
$\sqrt{2}$ smaller compared to $\sigma_H$ \cite[]{wood05}. The convolution of both
the Lorentzian and Gaussian profiles leads to a Voigt profile. The
relative radial velocity between the interstellar medium and LSR J1835+3259
is 22.9 km s$^{-1}$ based on heliospheric velocities of the dwarf of
8.4 km s$^{-1}$ \cite[]{desh12}
and velocities of $-14.4$ km s$^{-1}$ of the
interstellar medium using the LISM kinetic
calculator\footnote{http://sredfield.web.wesleyan.edu/} 
\cite[]{redf08}.  The resultant \Ly profile was convolved with the
line spread function described in the STIS Instrument Handbook
\cite[]{rile17}. 
We find a good fit to the observations shown  
as black line in Figure
\ref{fig:LYM-INT} with a source Voigt profile with
amplitude 1.2 $\times 10^{-14}$ erg s$^{-1}$ cm$^{-2} $
\AA$^{-1}$ and an \ion{H}{1} column density of 1  $\times 10^{18}$ cm$^{-2} $.
We see that the red wing of the observations is well reproduced by the
fit. 
The
total reconstructed \Ly profile is shown as the blue dotted lines in 
Figure \ref{fig:LYM-INT}. The total integrated \Ly source flux is 7.0  
$\times 10^{-15}$ erg s$^{-1}$ cm$^{-2} $. We note that there is a
partial degeneracy in the reconstruction of the source profile because
different combinations of the main free parameters, i.e., interstellar column densities,
temperatures and source profile shapes,
can lead to similar transmission profiles agreeing with the noisy
observations. 
Exploring this parameter
space we find that an uncertainty on the order of a factor of two to three
remains in constraining the integrated \Ly source flux.

As a simple independent check of our
reconstructed total \Ly flux source flux,
we alternatively use the observed 
and reconstructed
\Ly profiles of
three  
M2 and later dwarfs studied by
\cite{fran13}.
The \Ly spectra of these  three dwarfs are
 red wing dominated similar to the spectrum of LSR J1835+3259.
The dwarfs GJ 581 (M2.5), GJ 876 (M4),
and GJ 436 (M3) have fluxes at 1216.0 \A 
of $\sim$8, $\sim$9 and $\sim$9 $\times 10^{-14}$ erg s$^{-1}$ cm$^{-2} $
\AA$^{-1}$, respectively.
 The total reconstructed \Ly emission  from
these stars by \cite{fran13} is 3.0, 4.4 and 3.5  $\times 10^{-13}$ erg s$^{-1}$
cm$^{-2} $, respectively. This results in a ratio of $\sim$4 between
the total \Ly flux and the flux per \A at 1216 \AA. 
 At LSR J1835+3259
we measure a spectral flux of approximately 3 $ \times 10^{-15}$ erg s$^{-1}$
cm$^{-2} $ at 1216.0 \AA. Extending the scaling from the three M-dwarfs
to LSR J1835+3259, we expect a total \Ly flux of  1.2  $\times 10^{-14}$ erg s$^{-1}$
cm$^{-2} $, which is within 40\% of the modeled value of the previous paragraph.
The signal to noise of our observed red wing flux is five. But  due to
the partial degeneracy  in the reconstruction
processes of the the total source flux, the total \Ly flux should be
considered reasonable within a factor of two or three.

\section{IMPLICATIONS} \label{sec:implications}

In this section we compare the observed UV spectrum of LSR J1835+3259
with the spectrum of Jupiter's auroral emission and the UV
spectra of three low mass stars of spectral type M4.5 and later. 

\subsection{Comparison with Jupiter's auroral emission}
\begin{figure}
\plotone{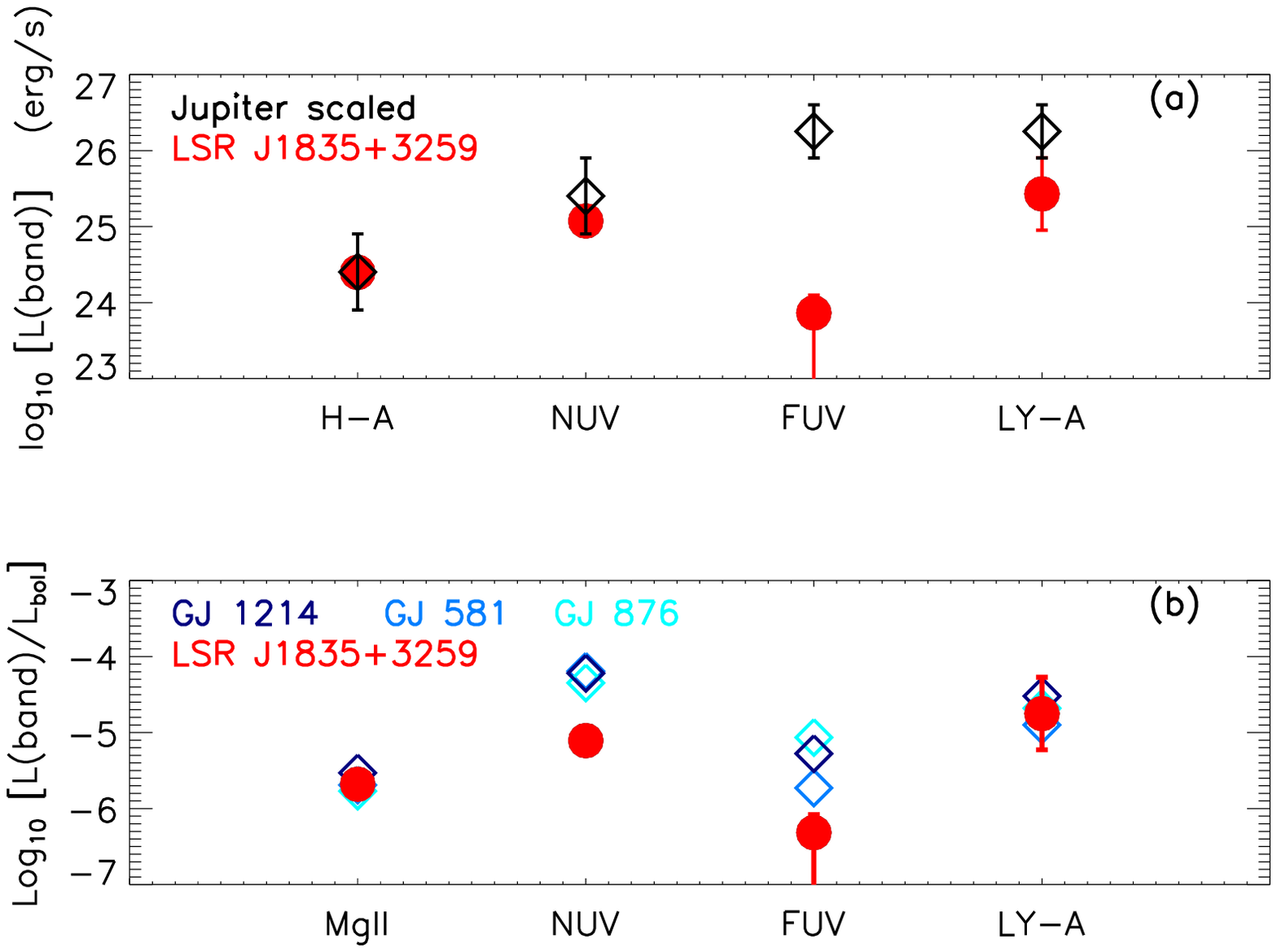}
\caption{Luminosities of LSR J1835+3259 for selected wavelength
  ranges in comparison with Jupiter in panel (a) and with three late M-dwarf
  stars in panel (b). \ion{Mg}{2} is excluded from the NUV band and
 \Ly is excluded from the FUV band in both panels.
Panel (a):  Jupiter is scaled to LSR J1835+3259 at
 H-$\alpha$ wavelength.  Bottom Panel (b): LSR J1835+3259 compared with
  GJ 581, GJ 876, and GJ 1214 but scaled to their respective
  bolometric luminosities (see also Table \ref{tab:energy_vgl}).
Error bars on the luminosities of LSR J1835+3259 include statistical
errors and the uncertainty from reconstructing the \Ly source
flux. When error bars are not visible, they are smaller than the used
symbols (for exact values see Table \ref{tab:energy_vgl}). Error bars on the values for Jupiter do not represent
measurement errors, but observed temporal-variability (see text and
table \ref{tab:energy_vgl} ). Measurement errors for the M-dwarfs  are
not explicitly provided in table 3 of \cite{fran16}, but should be 
30\% or smaller based on \cite{fran13}. 
\label{fig:energy_vgl}}
\end{figure}
\begin{deluxetable*}{llCCCCC}[b!]
\tablecaption{
 Luminosities
 in 
erg s$^{-1}$ 
for different wavelength ranges: LSR
  J1835+3259, Jupiter and mid/late-type M-dwarf stars compared. 
(1): \cite{hall15}, (2):
  \cite{bhar00}, (3):  \cite{pryo98}, (4): \cite{broa81a,glad02}, (5):
Based on the uncertainty to reconstruct the \Ly source flux
 (see Section \ref{subsec:Ly}),
 (6): \cite{fran16}.
Value ranges for
    Jupiter represent observed time-variability.
  \label{tab:energy_vgl}}

\tablecolumns{5}
\tablenum{4}
\tablewidth{0pt}
\tablehead{
\colhead{} 
&\colhead{Type}
&\colhead{H-$\alpha$}
&\colhead{NUV without \ion{Mg}{2}}
&\colhead{ \ion{Mg}{2}}
& \colhead{FUV without \Ly}
& \colhead{\Ly}
}
\startdata
LSR J1835+3259&M8.5& $ 2.5 \times 10^{24}$ (1) &  $ (1.2 \pm 0.2) \times
10^{25}$ & $ (3.2 \pm 0.2) \times
10^{24}$ &$  \le 7.3\times 10^{23}$ &  $ (2.7^{+5.4}_{-1.8}) \times 10^{25}$ (5) \\
Jupiter   & --&1-10 $\times 10^{17}$ (2)&1-10 $\times 10^{18}$ (3) &- &$ 1-5
\times 10^{19}$ (4) & $ 1-5 \times 10^{19}$ (2) \\
GJ 581 (6)   & M5 & - & 2.8 $\times 10^{27}$  &8.9 $\times 10^{25}$ & $ 8.1\times 10^{25}$  & $ 5.5 \times 10^{26}$  \\
GJ 876 (6)   & M5 &- & 2.2 $\times 10^{27}$  &8.3 $\times 10^{25}$ & $ 4.2 \times 10^{26}$  & $ 1.0 \times 10^{27}$  \\
GJ 1214 (6) &M4.5V& - & 8.5 $\times 10^{26}$  &4.2 $\times 10^{25}$ & $ 7.5\times 10^{25}$  & $ 4.3 \times 10^{26}$  \\
\enddata
\end{deluxetable*}

The Balmer H-$\alpha$ emission and emission at radio wavelengths from
LSR J1835+3259 have been interpreted by \cite{hall15} as 
auroral emission caused by electron beams originating within
the magnetosphere of LSR J1835+3259. 
 The emission of the dwarf could 
therefore be an upscaled version of the auroral emission of Jupiter. To test this
hypothesis we compare 
luminosities
 at optical and UV wavelengths of
LSR J1835+3259 with those of Jupiter in Figure \ref{fig:energy_vgl}(a) and
in Table \ref{tab:energy_vgl}. 
The black diamonds display the measured 
luminosities
 of Jupiter's
aurora at H-$\alpha$, the NUV, FUV, and \Lyy. 
In our comparison, we show the luminosities of \ion{Mg}{2} and
  \Lyy, separately, and therefore exclude their emission in the
  displayed NUV
  and FUV bands, respectively.
We multiplied the Jovian
luminosities
 by a factor of $\sim$10$^7$ for a better comparison with those of LSR
J1835+3259. In that way 
luminosities
 at the H-$\alpha$ wavelength are comparable for
Jupiter and LSR J1835+3259 in Figure \ref{fig:energy_vgl}(a).
The observed variability of Jupiter's emission at the various
wavelength ranges listed in Table 
\ref{tab:energy_vgl} is indicated as error bars in Figure 
\ref{fig:energy_vgl}(a). The uncertainties of the luminosities of LSR
J1835+3259 derived in the previous section are included
as well.

The Jovian 
 luminosities
 increase 
roughly by a factor of 10 from H-$\alpha$ to the NUV with most
of the emission in the NUV stemming from hydrogen. The auroral
luminosity
increases again roughly by a factor of 10 to the FUV, where the power
is split almost evenly between the \Ly emission and other wavelengths in the FUV. 
The
luminosity
 of LSR increases by a factor of five from the
H-$\alpha$ emission to the NUV. Roughly 1/4 of the NUV stems from the
\ion{Mg}{2} emission, the
other emission is uncertain. It could be partially due to TiO, but could contain
contributions, e.g.,  from Fe II lines as seen in M-dwarfs
\cite[e.g.,][]{fran13} or from hydrogen. In the FUV between 1330 and
1710 \A we find values of
$7.3 \pm 5.2 \times 10^{23}$  
erg s$^{-1}$.
The flux is very weak and at the detection
threshold. The derived FUV 
luminosities
 therefore should be
considered upper limits. Based on the assumption that LSR J1835+3259 scales similarly
to Jupiter throughout the optical and UV,  the dwarf, however, fails by 
a factor of approximately 1000 to reproduce Jupiter in the FUV wavelength
range without \Lyy. At \Ly wavelength, LSR J1835+3259 scales about a
factor of 10 weaker compared  to Jupiter.

Jupiter's UV spectrum contains the \ion{Mg}{2} doublet at 2800 \AA, but it
originates from reflectance of the solar \ion{Mg}{2} emission.  To the
authors' knowledge, \ion{Mg}{2} in emission from Jupiter was only seen in 
connection with the collision of comet Shoemaker-Levy 9.  After the
collision of the comet's G fragment,
emission from Jupiter's stratosphere 
 was reported by \cite{noll95a} and an outburst 
of \ion{Mg}{2} was observed before collision when the comet was inside
Jupiter's magnetosphere
\cite[]{feld96}. Due to absence of observed \ion{Mg}{2} emission
intrinsic to Jupiter we do not include \ion{Mg}{2} in the comparison
in Figure \ref{fig:energy_vgl}(a).

\subsection{Comparison with M-dwarfs}
In Figure \ref{fig:energy_vgl} (b), we
compare the
 luminosity
 of LSR J1835+3259 
with those of the three dwarf stars GJ
876, GJ 581 and GJ 1214, which are of spectral type M4.5 and
later. For these stars UV spectra have been previously obtained
with HST \cite[]{fran13,fran16,youn16,youn17,loyd16}.  
We compare the targets
normalized to their bolometric luminosities, respectively (see
\cite{fran16}). 
 Due to the absence of H-$\alpha$ in
 emission from these dwarfs, we do not include H-$\alpha$ in
 panel (b) of Figure \ref{fig:energy_vgl}.  Because we display the
 luminosities of \ion{Mg}{2} and \Lyy, separately, we exclude the
 luminosities of these two wavelengths in the NUV and FUV bands,
 respectively.

From the absolute values of
the luminosities listed in Table
\ref{tab:energy_vgl}, we find that LSR J1835+3259 is about a
factor of 200 less luminous in the NUV range and about a factor of 30
less luminous 
at \Ly and \ion{Mg}{2} wavelengths compared to the M5 dwarfs GJ 581 and GJ 876.
However, taking the luminosity of LSR J1835+3259 and the three M-dwarfs with
respect to their individual bolometric luminosities, we find that the line
luminosities of \ion{Mg}{2} and \Ly scale very similarly for all four objects.
  The \Ly to \ion{Mg}{2} 
luminosity
 ratio for LSR J1835+3259 is 10 with a 
statistical
uncertainty of $\pm$2 (plus the additional systematic uncertainty
 from the reconstruction of the \Ly source profile discussed
in Section \ref{subsec:Ly}). \cite{fran13} derive a similar 
 ratio of 10 $\pm$ 3 between the \Ly 
luminosity
 and the
luminosity
 of the \ion{Mg}{2} doublet for M-dwarfs (including the three M-dwarfs
 considered here).
 Comparable ratios also follow
from the scaling laws derived by \cite{shko14}.

For LSR J1835+3259,  the 
 \Ly 
luminosity
 comprises
about 70\%, i.e., a large fraction, of its total UV 
luminosity.
 This ratio is within the
range of 37\% to 75\% for M-dwarfs derived by \cite{fran13}. 
The relative NUV luminosities of the three M-dwarfs (excluding \ion{Mg}{2}) are slightly less than
a factor of 10 smaller compared to LSR J1835+3259. The relative FUV
luminosity (without \Lyy) of LSR J1835+3259 are a factor of 7 to 50 smaller compared
to those of the M-dwarfs. 

In addition to the mid-M-dwarfs of the previous section, it is
also interesting to compare the late-type M8 dwarf-star  Trappist-1 with
the M8.5 dwarf LSR J1835+3259, which are both of very similar
spectral type.
The \Ly 
luminosity
 of Trappist-1 is 1.4 $\times
10^{26}$
erg s$^{-1}$
as recently determined by \cite{bour17}. Trappist-1 is therefore approximately a
factor of three brighter at \Ly wavelength compared to 
LSR J1835+3259. Trappist-1 thus lies in its absolute \Ly luminosity
in-between LSR J1835+3259 and the discussed mid-dwarfs, which are a factor of 30
brighter. These observations thus confirm the trend of
decreasing \Ly 
luminosity
 with
increasing spectral type.

Summarizing the main findings of this subsection we see that LSR J1835+3259 is very similar to
mid-M-dwarfs in their \ion{Mg}{2} and \Ly luminosities when normalized to
their respective bolometric luminosity. We also find that the \Ly
 luminosity dominates the total FUV luminosity, both, in case of  LSR
J1835+3259 and in case of the three M-dwarfs.

\section{Conclusions and Discussion}
\label{sec:conclusions}

In the previous section we compared the UV spectrum of LSR J1835+3259
with the auroral UV spectrum of Jupiter and the spectra of mid/late
M-dwarf stars. 
We find that the observed 
luminosities
of LSR J1835+3259 
 in the UV 
are not consistent with 
an 
auroral spectrum expected from Jupiter due to the factor of 1000 discrepancy at FUV
wavelengths (without \Lyy). In addition, the emission at \Ly is lower by
approximately a factor of 10 compared to the averaged relative \Ly luminosity of
Jupiter. The error bars at \Ly wavelength however marginally overlap, where in the case of Jupiter the error
bars do not represent measurement uncertainties but observed
time-variability.

The emission of LSR J1835+3259 
resembles those of late-type M-stars  
 very well when considering the \ion{Mg}{2} to \Ly ratio. 
We also see that the
FUV luminosities (without \Lyy) is smaller than the \Ly luminosity for
LSR J1835+3259 as well as for the three $\sim$M5 dwarfs, where UV spectra are
available \cite[]{fran13,fran16,youn16,loyd16}. This is not the case for
Jupiter, where there is an approximate equipartition in \Ly and the FUV
band (without \Lyy) \cite[]{bhar00}.

The luminosity in the NUV and FUV bands of LSR J1835+3259 (without \ion{Mg}{2}
and  \Lyy, respectively) is, however, about an order of magnitude smaller compared to the three
mid-type M-dwarfs 
considered here.  
In case of
these  M-dwarf stars emission from \ion{C}{2}, \ion{Si}{4}, \ion{C}{4}, and
\ion{C}{1} is detected in their FUV spectra and \ion{Fe}{2} lines in
their NUV spectra \cite[]{fran13}.  These lines could not be
identified in the spectrum of LSR J1835+3259 even though they might
contribute to the observed fluxes in these wavelength ranges. 
The reason for
the weakness of these lines in LSR J1835+3259 is unclear. It could  
be a property of the later spectral type and its resultant
chromospheric structure, i.e. M8.5,  of LSR J1835+3259
 compared to the M4.5 and M5 dwarfs. It might be alternatively caused
 by different chromospheric flare activity levels. For LSR J1835+3259
 flare activity is observed at H-$\alpha$, while the three M-dwarf
 stars display weak chromospheric activity and
 H-$\alpha$ is in absorption \cite[]{gizi02,berg08,hall15,fran13,fran16,rein18}.

Based on this comparison the overall impression is that the dwarf LSR
J1835+3259 exhibits 
radiative properties which are generally more similar to those of low mass stars than  that of 
massive planets with externally driven auroral emission. 
A significant part of the
emission from LSR J1835+3259 might therefore be
 generated by internal
processes. Such processes could be chromospheric and coronal heating
driven by intrinsic magnetic activity, which results in
reconnection or wave heating
\cite[e.g.,][]{kuzm17,berd17}. 

It is worthwhile to point out that the heating and emission from quiescent
chromospheres is in general not a well understood process. In particular the ratio of the
two primarily discussed processes, i.e., mini-flares or wave-particle interaction due
to turbulent plasma waves, is being debated in the literature
\cite[e.g.,][]{gude97,gude03,aira98}. But even observational evidence for electron
beam generated emission from ultracool dwarfs might not uniquely
demonstrate the existence of
auroral processes caused by external power generators as in case of
the planets in the solar system. 
For example, observed emissions of the
transition region and  chromosphere of  the sun provide evidence of
heating by electron beams generated by nano-flare events
\cite[e.g.,][]{test14,reep15,dudi17}.


Even though our observations of LSR J1835+3259 are not consistent with
auroral activity similar to those on Jupiter, we
cannot entirely rule out that the emission of this dwarf contains
auroral emission, i.e.,  emission driven by electron or ion beams generated
within the magnetosphere of the dwarf. 
The space and plasma environment around ultracool dwarfs is not
sufficiently well understood. Therefore possible electron energization
mechanisms, resulting electron energy distributions, locations of
electron energy deposition within the chromosphere and the
resultant expected emission spectra throughout the electromagnetic
wavelengths range
are not well constrained.
The nature and occurrence of
 UV emission and aurora from ultracool dwarfs is thus a complex topic
because the emission is controlled by various physical parameters. The emission 
depends on intrinsic parameters such as their magnetic fields and atmospheres, but
it might also depend on unknown external
generators, which could power aurora.

\acknowledgments

This work is based on observations with the NASA/ESA Hubble Space
Telescope obtained at the Space Telescope Science Institute, which
is operated by the Association of Universities for Research in
Astronomy (AURA), Inc., under NASA contract NAS 5-26555. We thank J. Debes
for helpful comments on the scheduling and data processing of the
observations. CF and JS acknowledge funding by Verbundforschung f\"ur
Astronomie und Astrophysik through grant number 50 OR 170.
The work at Johns Hopkins University was supported by NASA through
grant HST-GO-14617.002-A from the Space Telescope Science Institute.




\end{document}